\RequirePackage{fix-cm}
\documentclass[smallextended]{svjour3}       % onecolumn (second format)
\smartqed  % flush right qed marks, e.g. at end of proof
\usepackage{graphicx}
\def\gsim{ \lower .75ex \hbox{$\sim$} \llap{\raise .27ex \hbox{$>$}} }
\def\lsim{ \lower .75ex\hbox{$\sim$} \llap{\raise .27ex \hbox{$<$}} }
\def\sc{Schwarzschild}

\begin{document}

\title{Blazar jets as the most efficient persistent engines%\thanks{Grants or other notes
}
% \subtitle{Do you have a subtitle?\\ If so, write it here}


\author{Gabriele Ghisellini         
%\and        Second Author %etc.
}


\institute{G. Ghisellini \at
              INAF -- Osservatorio Astronomico di Brera \\
%              Tel.: +123-45-678910\\
              \email{gabriele.ghisellini@inaf.it}           %  \\
}

\date{Received: date / Accepted: date}

\maketitle

\begin{abstract}

We have not identified for sure what is the mechanism launching, accelerating and collimating
relativistic jets.
The two most likely possibilities are the gravitational energy of the accreting 
matter or the rotational energy of a spinning black hole.
Even the evaluation of the jet power is not trivial, since the radiation from the jet is 
enhanced by relativistic beaming, and there are fundamental uncertainties concerning
the matter content of the jet (electron--proton or electron--positron plasma).
However, in recent years, there have been crucial advances mainly driven by the
richness of data in the $\gamma$--ray band. 
This is the band where blazars emit most of their electromagnetic power.
Furthermore, there are now large sample of $\gamma$--ray loud blazars 
covered by optical spectroscopy.
For the blazar sub--class of flat spectrum radio quasars (FSRQ) these data
provide measurements of the main emission lines and of the underlying continuum.
From these data, it is relatively easy to infer the bolometric luminosity of the
accretion disk.
The relativistic jet emission on one hand, and the disk luminosity on the other hand,
allows us to compare the jet power and the accretion luminosity.
Although the inferred jet power is subject to a few assumptions and
is somewhat model--dependent, 
it is possible to derive a lower limit to the jet power that is assumption--free and
model--independent.
Since this lower limit is of the order of the accretion luminosity, we infer that
the true jet power is larger.

\keywords{Relativistic jets \and BL Lac objects \and galaxies: quasars: general \and Radiative processes: general }
\end{abstract}

\section{Introduction}
\label{intro}
BL Lac objects and flat spectrum radio quasars (FSRQs) form the class
of blazars, radio--loud active galactic nuclei (AGN) whose jet is pointing toward us
(for recent reviews, see e.g. \cite{bottcher07}, \cite{gg11}, \cite{dermer14}.
FSRQs are powerful sources, with broad emission lines, contrary to 
BL Lacs, that are are less powerful and often line--less.
The observational distinction among them is in fact the equivalent width (EW) of their
broad emission lines: BL Lacs have EW$<$5\AA\ (rest frame, \cite{urry95}).
This divide is purely observational: 
objects that do have strong emitting lines could then be classified as
BL Lacs if their jet emission is particularly enhanced 
(see \cite{gg11}, \cite{sbarrato12} for an alternative definition).

Blazars emit most of their electromagnetic power by the synchrotron and
the inverse Compton processes. Their spectral energy distribution (SED)
is in fact characterized by two broad humps. 
The SED follows a sequence controlled by the bolometric luminosity:
when this is increasing, the peaks of the two main
radiation mechanisms shift to smaller frequencies, and the inverse Compton
component becomes more dominant \cite{fossati98}, \cite{donato01}, \cite{gg17}.

At the largest luminosities, the synchrotron peaks in the IR-sub--mm band,
and the slope above it is as steep as the slope in the $\gamma$--ray band
covered by {\it AGILE} and {\it Fermi}/LAT.
This makes the accretion disk clearly visible, and, at high redshifts, the observed 
accretion disk peak falls in the optical band.
In these cases we can directly measure its total luminosity. 
Otherwise, we have to rely on the observed broad emission lines.
Phenomenological relations between their relative weight \cite{francis91}, \cite{vanderberk01}
allows to calculate the total luminosity emitted by the broad lines.
Using an average covering factor of $\sim$10 we have an estimate of the accretion disk luminosity.

In addition, from the broad line FWHM and the continuum we can estimate the black hole mass.
This is the so called virial method, which relies on empirical correlations between the
disk luminosity and the distance of the BLR from the central engine.
The uncertainties of the method are quite large: a factor between 3 and 4 \cite{park12}, 
and they do not depend on the quality of data, but upon the dispersion of the empirical 
relations used.
Alternatively, we can directly fit the spectrum with a disk model, such as the
Shakura \& Sunjaev optically thick and geometrically thin accretion disk model 
\cite{ss73}, \cite{calderone13}. 
This simple model assumes zero black hole spin, and does not take into account
relativistic effects. 
Kerr models including all effects have been studied by, among others, \cite{li05} for binaries, 
and extended by \cite{campitiello18} to AGNs.

Blazars are among the  strongest $\gamma$--ray emitting sources.
The third catalog of AGNs detected by {\it Fermi}/LAT lists $\sim$1,500 sources,
of which the vast majority are blazars \cite{3lac15}.
This made possible to make population studies, such as the $\gamma$--ray blazar 
luminosity function and their contribution to the $\gamma$--ray background
\cite{ajello12}, \cite{ajello14}.

\begin{figure}
  \includegraphics[width=12cm]{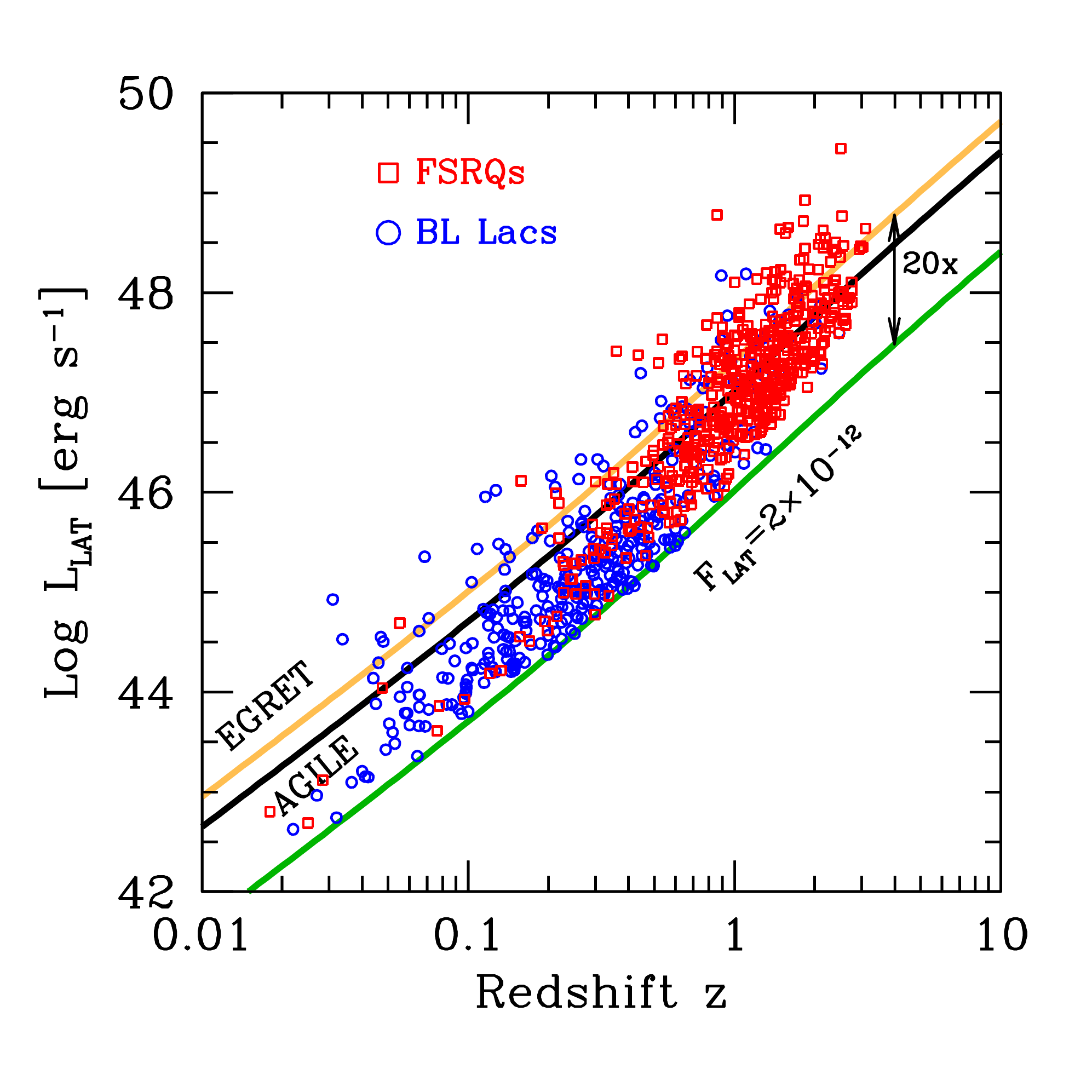}
\vskip -0.7 cm
\caption{The $\gamma$--ray luminosity of blazars detected by {\it FERMI}/LAT as a
function of their redshift. We also plot three solid curves, corresponding (roughly) to
the LAT, AGILE, and EGRET sensitivity assuming for simplicity
that the $\gamma$--ray spectrum is the same. 
}
\label{lg_z}       % Give a unique label
\end{figure}

Fig. \ref{lg_z} shows the $\gamma$--ray luminosities of BL Lacs and FSRQs as a function
of their redshift.
We have also drawn three lines indicating the approximate limiting sensitivity
of EGRET, AGILE and {\it Fermi}/LAT.
It is approximate because the same spectral index is used for all sources.
This figure shows the improvement that AGILE and especially LAT made possible
in our knowledge of blazars. 
It also makes clear that EGRET could explore only the tip of the iceberg
of the entire blazar population. 
Furthermore, since all blazars vary greatly especially in $\gamma$--rays,
EGRET probably detected most blazars only when they are in high state.
LAT returns instead a more balanced view.

%-----------------------------------------------

\section{Comparing jet power and accretion luminosity}

The first attempt to derive the relation between the jet and the accretion disk power
was made by Rawling \& Saunders in 1991 \cite{rawlings91}.
They estimated the average jet power considering the extended
radio emission, calculating their minimum energy through equipartition considerations, 
and estimating their lifetime considering the time needed to expand and the presence of 
a cooling break in the spectrum.  
Then they collected for all the sources the luminosity of the narrow lines.
They found a strong correlations between jet power and line luminosity, but  this, per se,
is not unexpected, since both quantities depend on distance.
What is interesting was that the narrow line luminosity was always a factor 100 less that the
jet power.
Since the covering factor of the narrow lines is $\sim 10^{-2}$, they concluded that the average jet power
energizing the lobes and the accretion luminosity were of the same order.

Stimulated by this result, Celotti \& Fabian \cite{celotti93} tried to improve the estimate of the jet power
by considering the jet emission at the parsec (e.g. VLBI) scale to find the number of particles needed
to account for the emission and the Lorentz factor needed not to overproduce the X--ray emission
through the synchrotron--self--Compton (SSC) process.
They confirmed the earlier results.
Later, the same was performed using broad line luminosities instead of the narrow ones \cite{celotti97}.

\begin{figure*}
  \includegraphics[width=1.\textwidth]{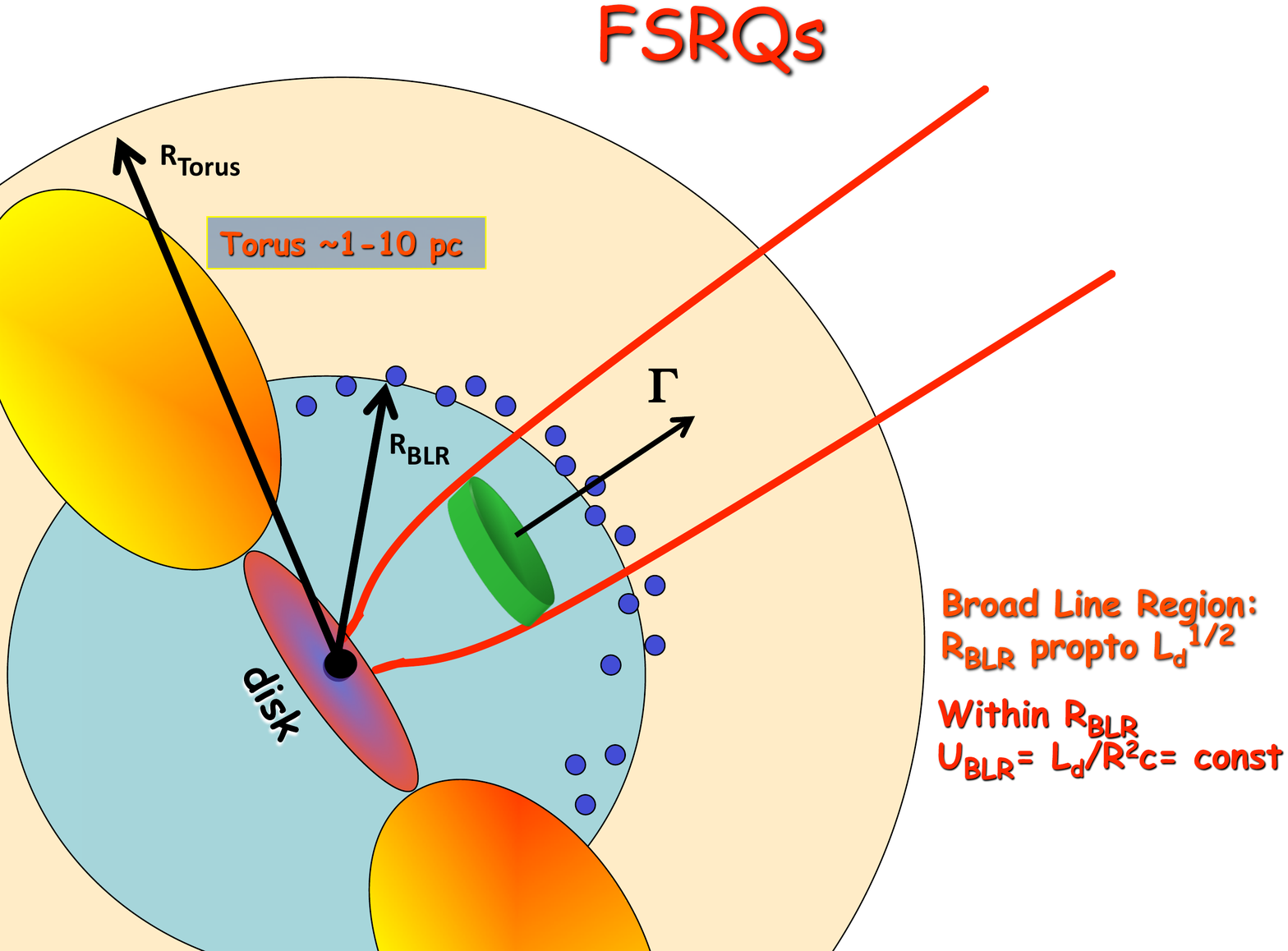}
\caption{Cartoon of the assumed geometry for the model,
see \cite{gg09}. At some distance $R_{\rm diss}$ from the black hole, 
the jet produces most of its radiation. 
This region can be inside the BLR or within $R_{\rm torus}$,
the dimension of the molecular torus. 
The radiation energy densities within the BLR and the torus
are constant,
since both $R_{\rm BLR}$ and $R_{\rm torus}$ scale as $L_{\rm d}^{1/2}$.
This structure is valid for FSRQs only, since BL Lacs have a radiatively
inefficient disk, unable to photo--ionize the BLR.
}
\label{fsrq}       % Give a unique label
\end{figure*}

At the same time of these pioneering studies, EGRET discovered that blazars are very strong $\gamma$--ray emitters,
and that this emission is the most variable.
The first models thought that the $\gamma$--rays were SSC emission 
\cite{maraschi92}, \cite{bloom92} \cite{bloom93}. 
Soon it became clear that variability at different frequencies was correlated, strongly suggesting 
that the same population of particles, in only one region, was responsible for the emission.
This was the birth of the leptonic, one--zone model. 
The location of this one--zone is still controversial:
early suggestions located it close to the accretion disk \cite{dermer92},
or within the BLR \cite{sikora94}, or beyond the BLR but within the molecular torus size
\cite{sikora02}.
In these models, the seed photons for the inverse Compton process are produced externally
to the jet (External Compton, EC for short).
This process is highly efficient, since the radiation energy density, as seen
in the comoving frame, is enhanced by $\Gamma^2$.
In general, we have two constraints that limits the location of the emitting region:
i) if too close to the disk, the produced $\gamma$--rays are absorbed in $\gamma$--$\gamma$ collisions,
and ii) if very distant from the black hole, the dimension of the emitting region
becomes too large to account for the observed short variability.
Distances of the order of $\sim$10$^3$ \sc\ radii are suggested.
See the cartoon in Fig. \ref{fsrq}.

The one--zone model had the advantage to greatly simplify the modelling while
reducing the number of free parameters. 
For low power blazars, lacking broad lines and thermal components, the SSC
model is still the leading model, and has the virtue of having no degeneracies:
given some basic observables, the solution is unique \cite{tavecchio98}.
EC models, on the other hand, can offer a unique solution if some extra information
is supplied or can be derived from the fitting, such as the black hole mass and the disk luminosity
(see \cite{gg09}).

%-------------------------------------------------

\section{Electrons and positrons}

If we calculate the jet power estimating how many leptons are necessary
to produce the received flux we face a crucial problem.
Any particle distribution fitting the data has to be relatively
steep: $N(\gamma)\propto \gamma^{-p}$, with $p\ge 2$,
so that the total number of the emitting particles depends strongly on
the minimum particle energy, or their minimum random Lorentz factor $\gamma_{\rm min}$.\
The synchrotron emission is self--absorbed for values of $\gamma \lsim$50--100
and the SSC flux is always hidden by the synchrotron one
at low energies.
In these condition we cannot estimate $\gamma_{\rm min}$, that corresponds
to an uncertainty on the jet power of the order of $\sim \gamma_{\rm min}^{p-1}$.
But for powerful blazars the soft X--ray emission is produced by the EC process.
In this case we do see the effects of changing $\gamma_{\rm min}$
since the soft X--ray shape should become harder below 
$\nu_{\rm Ly\alpha} \gamma^2_{\rm min}\Gamma^2 \sim$ 1 keV $\times \gamma^2_{\rm min}(\Gamma/10)^2$.
Usually, we do not see a break in the soft X--ray spectrum,
corresponding to $\gamma_{\rm min}<$ a few.
This agrees with the typical values estimated by the radiative cooling occurring
in the emitting region.

\begin{figure*}
  \includegraphics[width=1.\textwidth]{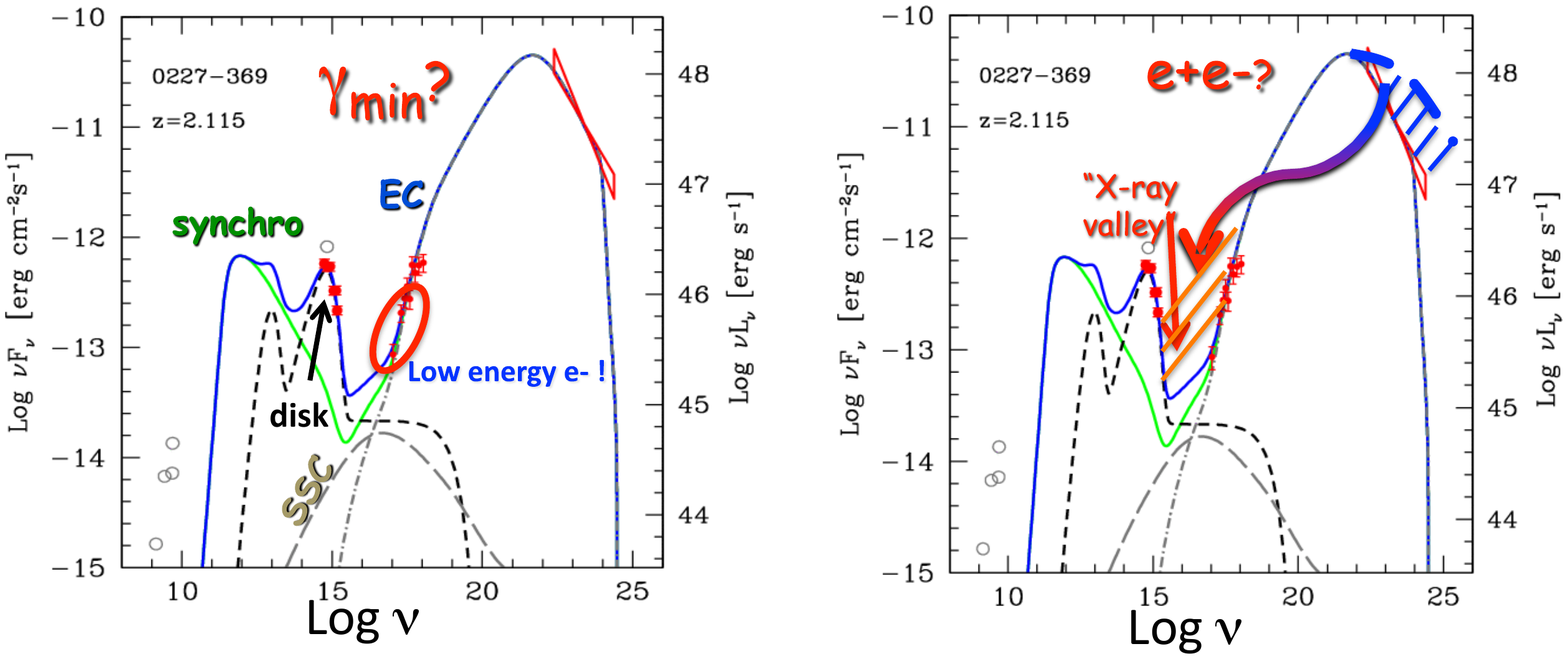}
%  \includegraphics[width=0.6\textwidth]{pairs.pdf}
\caption{In the past there were two major uncertainties concerning
the estimate of the jet power.
The first was the difficulty to estimate the lowest electron energies,
but now we do have information about it looking at the soft X--rays,
where, in the EC scenario, the lowest energy electron contribute to the emission.
The second is the importance of pairs. 
As illustrated in the right panel, if part of the $\gamma$--rays do form pairs, we expect 
that they emit in the X--ray band, filling the ``valley" between the synchrotron and th
Compton humps.}
\label{gmin}       % Give a unique label
\end{figure*}
%-------------------------------------------------

The other uncertainty concerns the presence of protons, or, equivalently,
the number of electron--positron pairs.
This is because, at least in powerful blazars, the bulk motion of protons,
even if cold in the comoving frame, dominates the jet power if we assume
that there is one proton for emitting electron.
If instead there are no protons, i.e. for a pure pair jet, the jet power
estimate is a factor $ \langle\gamma \rangle m_{\rm e} /m_{\rm p}\approx 10^{-2}$ smaller.
The presence of both protons and pairs corresponds to intermediate values of
the jet power.
The counter--arguments concerning the presence of e$^-$--$e^+$ pairs are:
\begin{itemize}
\item It is not easy to produce pairs in the first place.  
If we are using even a relatively small fraction of the $\gamma$--ray photons produced in the emitting region,
we are nevertheless using a large power. 
The produced pairs, that are born relativistic, emit efficiently, mostly in the X--rays
where the SED of powerful blazars has a ``valley" \cite{gg96}. See the right panel of Fig. \ref{gmin}.
We do not see any sign of this reprocessed emission.

\item If pairs are produced at the start, very close to the jet apex, there is a maximum number
of them that can survive.
In fact, if their optical Thomson optical depth is larger than unity and they are cold, they annihilate
efficiently until the optical depth is below unity.
If they are hot, annihilation is reduced, but they are bound to cool in a short time.

\item Pairs can be the result of hadronic cascades, but in this case we have to account for
the bulk motion of ultra--relativistic protons, and this increases the power demand.

\item In powerful blazars relativistic pairs efficiently cool on external radiation
and recoil, braking the jet \cite{sikora96}, \cite{ggtav10} if there are more than
$\sim$10--20 pairs per proton.

\item The total power in radiation (integrated over the entire solid angle)
often exceeds the power in the relativistic emitting
particles, because the radiative cooling time is shorter than the light crossing time. 
This requires continuous injection of new fresh particles or re--acceleration. 
If these processes are the result of the dissipation of a fraction of the 
jet kinetic or magnetic power, they must be larger than the radiated power. 
The jet Poynting flux is constrained by the observed synchrotron luminosity that is smaller
than the inverse Compton one. 
Bulk motion of protons is therefore required. 

\end{itemize}
We conclude that $e^-$--$e^+$ pairs can be present in jets, but their number is severely limited 
to a be a few per proton.
A slightly larger number (i.e. 15 per proton) has been invoked by \cite{sikora16}
to explain the different median power of a sample of FRII and FSRQs calculated
though radio--lobe calorimetry.
Such a number of pairs corresponds to a jet kinetic power of the same order of the radiated
luminosity, implying that jets are highly efficient radiators. 
Alternatively, radio--lobes may contain hot protons, not accounted for by calorimetry.

\section{Jet power}

Jets can carry energy in different forms:
\begin{itemize}
\item {\it Radiation ---} The observed emission has been produced in the jet, that
spends part of its power to produce it. 
This is not merely the luminosity measured in the comoving frame,
because it needs some extra power to move the emitting particles 
in order to blue--shift the photons we receive and in order to
account for the different rate of arrival.
Both these effects are measured by the relativistic Doppler factor
$\delta \equiv[\Gamma(1-\beta\cos\theta_{\rm v})]^{-1}$,  where $\theta_{\rm v}$ is the viewing angle.
The observed ($L$) and the comoving ($L^\prime$) jet bolometric luminosity are linked by $L =\delta^4 L^\prime$.
If the luminosity is isotropic in the comoving frame, as can be the case for synchrotron and SSC:
\begin{equation}
P_{\rm r} = 2\times  \int_{4\pi} {L^\prime\over 4\pi} \delta^4 d\Omega = 2 {L^\prime\over 2\Gamma^4} 
\int_{-1}^{1} {d\cos\theta \over (1-\beta\cos\theta)^4} = 2 {\Gamma^2 L\over \delta^4}  
\left(1+{\beta^2\over 3}\right)
\end{equation}
The factor 2 accounts for the presence of two jets.
If the radiation is not isotropic in the rest frame, as in the EC process, 
%Ghisellini \& Tavecchio
we have \cite{ggtav10}:
\begin{equation}
P_{\rm r} = 2\times {\langle L^\prime\rangle \over 4\pi} \int_{4\pi} {\delta^6\over \Gamma^2} d\Omega 
\sim {32 \over 5}  {\Gamma^4 \over \delta^6} L  
\end{equation}
When $\Gamma=\delta$, implying that the viewing angle $\theta_{\rm v}=1/\Gamma$, 
the two expressions differ by a factor $5/12$.
There is an alternative way to calculate $P_{\rm r}$, that is useful also for 
calculating the other the forms of carried energy.
It consists to calculate the flux across a cross section (of radius $r$) of the jet:
\begin{equation}
P_{\rm r} = 2\times \pi r^2 \Gamma^2 U^\prime_{\rm r}  c \sim \pi r^2  \Gamma^2 {L\over \delta^4 4\pi r^2 c} c
= {\Gamma^2\over 2 \delta^4}   L 
\end{equation} 
The reason of the $\Gamma^2$ term is due to the transformation of the energy density
from the comoving to the observer frame: one term for the blue--shift, the other term for
the different number density in the two frames.

\vskip 0.3 cm
\item {\it Magnetic field ---} The Poynting flux is 
\begin{equation}
P_{\rm B} = 2\times \pi r^2 \Gamma^2 U^\prime_B  c  
\end{equation} 
where $U^\prime_B\equiv (B^\prime)^2/(8\pi)$ and the factor 2 accounts again for the presence of two jets.
\vskip 0.3 cm
\item {\it Relativistic leptons ---} The power in the bulk motion of relativistic leptons is
\begin{equation}
P_{\rm e} = 2\times \pi r^2 \Gamma^2 U^\prime_{\rm e} \beta c  = 
2\pi r^2 \Gamma^2 \beta \, \langle \gamma\rangle \, m_{\rm e}c^3 \, n^\prime_{\rm e}
\end{equation} 
where $U^\prime_{\rm e} =m_{\rm e} c^2 \int_{\gamma_{\rm min}}^{\gamma_{\rm max}} N(\gamma)\gamma d\gamma
= \langle \gamma\rangle \, m_{\rm e}c^2 \, n^\prime_{\rm e}$.
Usually, $n_{\rm e}$ and $\langle \gamma\rangle$ are calculated through the properties
of the observed emission. 
There could be other leptons not participating to the emission and therefore 
not accounted for when we estimate the $N(\gamma)$ distribution.

\vskip 0.3 cm
\item {\it Protons ---} The power in the bulk motion of protons is
\begin{equation}
P_{\rm p} = P_{\rm e} \, {n_{\rm p}\over n_{\rm e} }\, 
{ \langle \gamma_{\rm p} \rangle m_{\rm p}\over \langle \gamma \rangle m_{\rm e}}
\end{equation} 
If protons are cold, then $\langle \gamma_{\rm p} \rangle=1$, but if there is a hot component,
possibly highly relativistic, then the power increases.
This is indeed a possibility, especially after the discovery of the likely association
of one high energy neutrino detected by Icecube with the BL Lac TXS 0506+056 \cite{icecube18}.
For the rest of this paper we will assume cold protons, and one proton for each emitting electron.
With these working assumptions, note that the presence of protons, even if cold, is very important 
when the $\langle \gamma \rangle$ of the leptons is small, namely for FSRQs. 
On the other hand, TeV emitting BL Lacs can have $\langle \gamma \rangle > m_{\rm p}/m_{\rm e}$.
In these sources, protons (if cold) do not contribute much to the total jet power.

\end{itemize}

\section{Results}

To compare the jet power with the accretion luminosity we need blazars
that have been detected in the $\gamma$--ray band and that have been observed spectroscopically, 
with detected broad emission lines.
The high energy emission allows us to know the entire bolometric luminosity,
the broad emission lines allows us to estimate the disk luminosity.

To this aim, we have used the sample of blazars studied by \cite{shaw12} (FSRQs) 
and \cite{shaw13} (BL Lacs). 
Only very few BL Lacs were selected, only those showing the presence of broad lines.
We  have collected for all blazars of the total sample the available 
archival data, and we have applied the one--zone leptonic model described in \cite{gg09}.
The results concerning the 217 blazars of this sample were presented
in \cite{gg14} and \cite{gg15}.

The SED coverage was much better than in the past because for all
blazars in the sample we had:
i) the spectroscopic data (that are used to find the disk luminosity
and the black hole mass), 
ii) the $\gamma$--ray luminosity (used to infer the jet radiated power),
iii) the far IR data given by {\it WISE} satellite (that help
to shape the jet+torus continuum), 
iv) the high frequency radio data from the {\it WMAP} and {\it Planck} satellites 
helping to define the synchrotron peak.

Then we have added  a sample of $z>2$ blazars detected by {\it Swift}/BAT \cite{gg10chasing},
a sample of $z>4$ blazars selected from \cite{sbarrato13} and
two powerful FSRQs detected by {\it NuSTAR} \cite{sbarrato16}.

\begin{figure*}
  \includegraphics[width=1.1\textwidth]{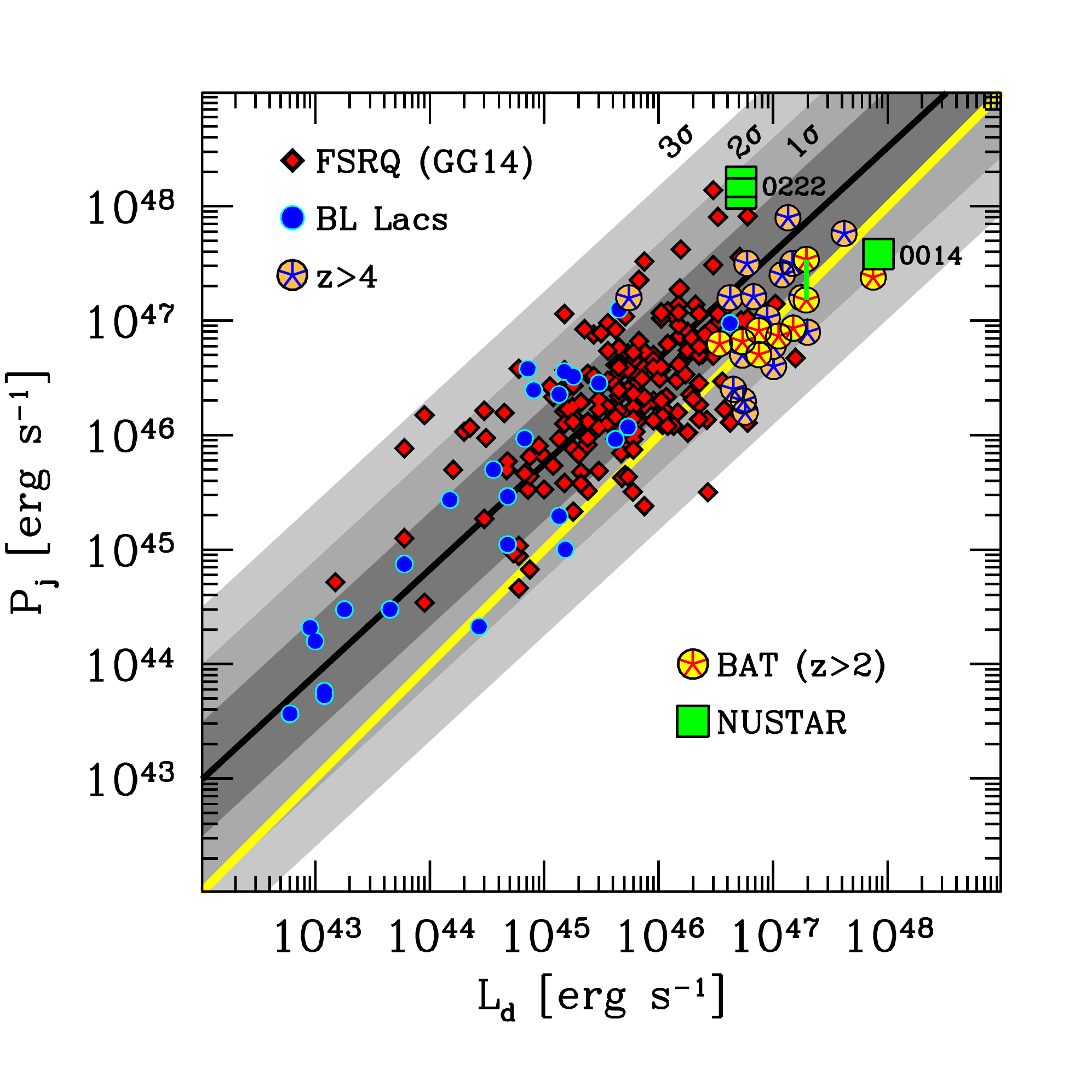}
\vskip -0.7 cm
\caption{The jet power as a function of the accretion luminosity
for different blazar samples.
Also the BL Lacs shown in this figures had some broad lines in their spectrum.
The sources belong to the sample studied by \cite{shaw12}, \cite{shaw13} and \cite{gg14};
two powerful FSRQs detected by {\it NuSTAR} \cite{sbarrato16};
a sample of $z>2$ blazars detected by {\it Swift}/BAT \cite{gg10chasing} and 
a sample of $z>4$ blazars selected from \cite{sbarrato13}.
The yellow line is the equality line, the black line is the best fit.
}
\label{pjld}       % Give a unique label
\end{figure*}
%-------------------------------------------------

We find $P_{\rm r} \sim L_{\rm d}$.
This results is model independent, since $P_{\rm r} \sim L/\Gamma^2$, and
while $\Gamma$ is indeed found through modelling, the typical values found agree 
with independent estimates given by superluminal motion and other beaming indicators
(such as the radio brightness temperature).

{\it This is a lower limit to the jet power and does not depend on the uncertainties
regarding the particle distribution and the importance of electron--positron pairs.
This result is solid.}

Then, to go from $P_{\rm r}$ to $P_{\rm j}= P_{\rm r} +P_{\rm e} +P_{\rm p} +P_{\rm B}$
we do have to apply a model. 
We used the model described in \cite{gg09}. 
Fig. \ref{pjld} shows $P_{\rm j}$ as a function of $L_{\rm d}$
assuming one cold proton per emitting electron.
This gives, on average, $P_{\rm j} \sim 10 L_{\rm d}$.

\section{Discussion}

How can the jet power be proportional to $L_{\rm d}$, but larger than it?
We have two possibilities.
The first, suggested by the fact that the two powers are proportional,
is that it is accretion that powers the jet. 
The gravitational energy dissipated by the accreting $\dot M$ 
is transformed into heat only partly while the rest goes to power the jet \cite{jolley08}, \cite{jolley09}.
The total accretion efficiency $\eta$ is used to power the jet with an efficiency $\eta_{\rm j}$,
and to heat the disk with an efficiency $\eta_{\rm d}$:
\begin{equation}
P_{\rm j} = \eta_{\rm j} \dot Mc^2; \,\,\,\,\, L_{\rm d}=\eta_{\rm d} \dot Mc^2;
\,\,\,\,\, \eta=\eta_{\rm j}+\eta_{\rm d}
\end{equation}
Then $P_{\rm j}= 10 L_{\rm d}$ requires $\eta_{\rm j}= 10 \eta_{\rm d}$.
This has an important consequence: if there is a jet, the disk efficiency can be smaller (even by a factor 10)
than what is foreseen for standard accretion models.
To produce a given $L_{\rm d}$, $\dot M$ must therefore be larger, and this helps
jetted sources to have black holes that grow faster (especially at large redshifts \cite{gg13}).

The second possibility is that jets are powered not by accretion, but by the rotational energy
of the black hole, by the Blandford-Znajek process \cite{blandford77}.
The total energy that can be extracted by a maximally spinning black hole 
is the 29\% of its mass, namely $5\times 10^{62}$ erg for a $10^9 M_\odot$ black hole.
This is enough to power a jet with $P_{\rm j}=10^{47}$ erg s$^{-1}$ for 166 million years.
As Cavaliere and D'Elia \cite{cavaliere02} and Ghisellini and Celotti \cite{gg02}
proposed in 2002, this can be the engine for powering blazars.
This has also been confirmed by numerical simulations \cite{tchekhovskoy12} finding
an average jet power corresponding to $\sim$1.5$ \dot M c^2$ that obviously 
indicates that the power of the jet does not come from direct accretion, but from the rotational
black hole energy (in turn originated by previous accretion, and now released in a
efficient way).

The fact that $P_{\rm j}\propto L_{\rm d}$ in this scenario can be explained if the magnetic
field energy density necessary to tap the rotational energy of the black hole is proportional
to $\dot M$ and hence to $L_{\rm d}$. 
This is possible if $B^2 \simeq \rho c^2$, where $\rho$ is the disk density.

Furthermore, when the accretion disk is radiatively efficient, it produces most of its luminosity
in the optical--UV, ionizing the gas of the BLR, that can indeed emit broad emission lines,
But al low rates of accretion, the disk becomes radiatively inefficient (ADAF
\cite{rees82}, \cite{narayan94};  ADIOS \cite{blandford99} and so on), 
and most of its (already reduced) luminosity is not emitted in the UV \cite{mahadevan97}.
The clouds of the BLR are not photo-ionized, and no broad line is produced.
This explains the divide between FSRQs and BL Lacs \cite{gg09divide} and possibly
their different cosmic evolution \cite{cavaliere02}.

%\begin{acknowledgements}
%\end{acknowledgements}



\end{document}